\documentclass[dvipsnames,format=sigconf,anonymous=false,review=false,hidelinks,natbib=false]{acmart}

\usepackage{url}
\usepackage{hyperref}
\usepackage[abbreviate=true, dateabbrev=true, isbn=false, doi=true, urldate=comp, url=false, maxbibnames=9, backref=false, backend=biber, style=numeric-comp, language=american, giveninits=true, sortcites=true, sorting=none]{biblatex}

\definecolor{aggreen}{rgb}{0.0, 0.8, 0.6}
\newcommand\ag[1]{\textcolor{black}{#1}}
\newcommand{\head}[1]{\par\noindent\textbf{#1:}\space}

\usepackage{graphicx}
\usepackage{caption}
\usepackage{subcaption}
\usepackage{multirow}
\usepackage{comment}
\usepackage{adjustbox}
\usepackage{enumitem}
\usepackage{fancybox}
\usepackage[framemethod=tikz]{mdframed}
\mdfdefinestyle{mystyle}{%
  rightline=true,
  nobreak=false,
  innerleftmargin=5,
  innerrightmargin=5,
  outerlinewidth=1pt,
  topline=false,
  leftline=true,
  bottomline=false,
  skipabove=\topsep,
  skipbelow=\topsep
}
\newcommand{\method}[0]{SEIDR}
\newcommand{\synthesize}[0]{SYNTHESIZE}
\newcommand{\execute}[0]{EXECUTE}
\newcommand{\instruct}[0]{INSTRUCT}
\newcommand{\instructs}[0]{INSTRUCT$^{\text{static}}$}
\newcommand{\instructllm}[0]{INSTRUCT$^{\text{LLM}}$}
\newcommand{\debug}[0]{DEBUG}
\newcommand{\rank}[0]{RANK}
\newcommand{\beamwidth}[0]{W}
\newcommand{\treearity}[0]{N}
\newcommand{\expectedoutput}[0]{O}
\newcommand{\synthmodel}[0]{$p_\text{synth}(\text{input}, \text{instr})$}
\newcommand{\debugmodel}[0]{$p_\text{debug}(\text{input}, \text{instr})$}
\newcommand{\textmodel}[0]{$p_\text{text}(\text{input})$}

\DeclareRobustCommand{\citex}[1]{\citeauthor{#1}~\cite{#1}}

\DeclareSourcemap{
  \maps[datatype=bibtex]{
    \map{
      \pertype{misc}
        \step[fieldsource=eprint, match={arXiv:}, replace={}]
        \step[fieldset=DOI, null]
    }
  }
}

\AtBeginDocument{%
  \providecommand\BibTeX{{%
    \normalfont B\kern-0.5em{\scshape i\kern-0.25em b}\kern-0.8em\TeX}}}

\copyrightyear{2023}
\acmYear{2023}
\setcopyright{rightsretained}
\acmConference[GECCO '23]{Genetic and Evolutionary Computation
Conference}{July 15--19, 2023}{Lisbon, Portugal}
\acmBooktitle{Genetic and Evolutionary Computation Conference (GECCO '23),
July 15--19, 2023, Lisbon, Portugal}\acmDOI{10.1145/3583131.3590481}
\acmISBN{979-8-4007-0119-1/23/07}

\makeatletter
\def\@copyrightpermission{
  \hspace*{0mm}\includegraphics[width=2cm]{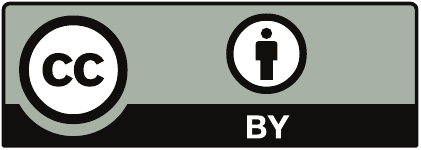}%
  \hspace*{2mm}\raisebox{2.5mm}[25pt][5pt]{%
          \parbox{\columnwidth}{\footnotesize This work is licensed under a Creative Commons \\ Attribution 4.0 International (CC BY 4.0) license.}%
  }%
}%
\makeatother

\begin{document}
\thispagestyle{plain}
\pagestyle{plain}

\settopmatter{authorsperrow=4}

\title{Fully Autonomous Programming with Large Language Models}

\author{Vadim Liventsev}
\email{v.liventsev@tue.nl}
\orcid{0000-0002-6670-6909}
\authornote{~Both authors contributed equally to this research.}
\authornote{~Corresponding author.}
\affiliation{%
  \institution{TU Eindhoven \& \\Philips Research}
  \country{The Netherlands}
}

\author{Anastasiia Grishina}
\email{anastasiia@simula.no}
\orcid{0000-0003-3139-0200}
\authornotemark[1]
\affiliation{%
  \institution{Simula \& University of \\Oslo} %
  \country{Norway}
}

\author{Aki H\"{a}rm\"{a}}
\email{aki.harma@philips.com}
\orcid{0000-0002-2966-3305}
\affiliation{%
  \institution{Philips Research}
  \city{Eindhoven}
  \country{\linebreak{}The Netherlands}}

\author{Leon Moonen}
\email{leon.moonen@computer.org} %
\orcid{0000-0002-1761-6771}
\affiliation{%
  \institution{Simula \& BI Norwegian Business School}
  \city{Oslo}
  \country{Norway}%
}

\renewcommand{\shortauthors}{Liventsev and Grishina, et al.}

\begin{abstract}

Current approaches to program synthesis with Large Language Models (LLMs) exhibit a ``near miss syndrome'': they tend to generate programs that semantically resemble the correct answer (as measured by text similarity metrics or human evaluation), but achieve a low or even zero accuracy as measured by unit tests due to small imperfections, such as the wrong input or output format. 
This calls for an approach known as Synthesize, Execute, Debug (SED), 
whereby a draft of the solution is generated first, followed by a program repair phase addressing the failed tests.
To effectively apply this approach to instruction-driven LLMs, one needs to determine which prompts perform best as instructions for LLMs, as well as strike a balance between repairing unsuccessful programs and replacing them with newly generated ones.
We explore these trade-offs empirically, comparing replace-focused, repair-focused, and hybrid debug strategies, as well as different template-based and model-based prompt-generation techniques.
We use OpenAI Codex as the LLM and Program Synthesis Benchmark 2 as a database of problem descriptions and tests for evaluation. 
The resulting
framework outperforms both conventional usage of Codex without
the repair phase and traditional genetic programming approaches.
 \end{abstract}



\maketitle

\section{Introduction}
\label{sec:intro}

Automatic programming has been an important goal of the Artificial Intelligence field almost since its inception~\cite{manna1971:automatic}, promising to reduce the workload of software developers by automatically solving some of the tasks they face.
More recently, program synthesis has emerged as an interpretable alternative~\cite{bastani2022:interpretable} to black-box machine learning methods that lets human experts understand, validate and edit the algorithms generated by artificial intelligence.
In addition to the scientific benefits of such knowledge, it extends the benefits of machine learning to domains, such as embedded systems where it is technically challenging~\cite{dhar2021:survey} or healthcare where it is avoided for safety reasons~\cite{connolly2023:systematic,jia2022:role}.

The predominant methodology in automatic programming has shifted from deductive programming~\cite{manna1992:fundamentals,alur2015:syntaxguided} to genetic and evolutionary methods~\cite{ahvanooey2019:survey} to, more recently, large autoregressive language models trained on corpora of source code due to their remarkable capability for zero-shot generalization~\cite{chen2021:evaluating}.
However, even state-of-the-art models fine-tuned on a specific class of programming tasks still require a costly filtering step where the LLM outputs that do not compile or pass tests are discarded~\cite{li2022:competitionlevel}.
These outputs tend to be superficially similar to correct solutions~\cite{ren2020:codebleu} despite failing to produce the expected output, a phenomenon known as "near miss syndrome" or "last mile problem"~\cite{bavishi2022:neurosymbolic}. 

Given these challenges, research in machine learning on source code~\cite{allamanis2018:survey} tends to focus on restricted domain-specific languages~\cite{chen2021:latent,polozov2015:flashmeta,liventsev2021:bf} or automating specific parts\footnote{~similarly to autonomous driving~\cite{grigorescu2020:survey,marcano2020:review}} of the software development process~\cite{lu2021:codexglue,niu2023:crosscodebench} such as code search~\cite{husain2020:codesearchnet}, code translation~\cite{roziere2020:unsupervised}, detection of issues~\cite{fernandes2016:reviewbased,chakraborty2021:deep}, improvement~\cite{petke2018:genetic} and repair~\cite{legoues2019:automated} rather than fully autonomous programming in a programming language popular with human developers~\cite{:tiobe}.
However, two recent innovations potentially make the latter task tractable.

\begin{figure*}[t]
    \centering
    \includegraphics[width=\linewidth,trim={0mm 8mm 0mm 0mm}]{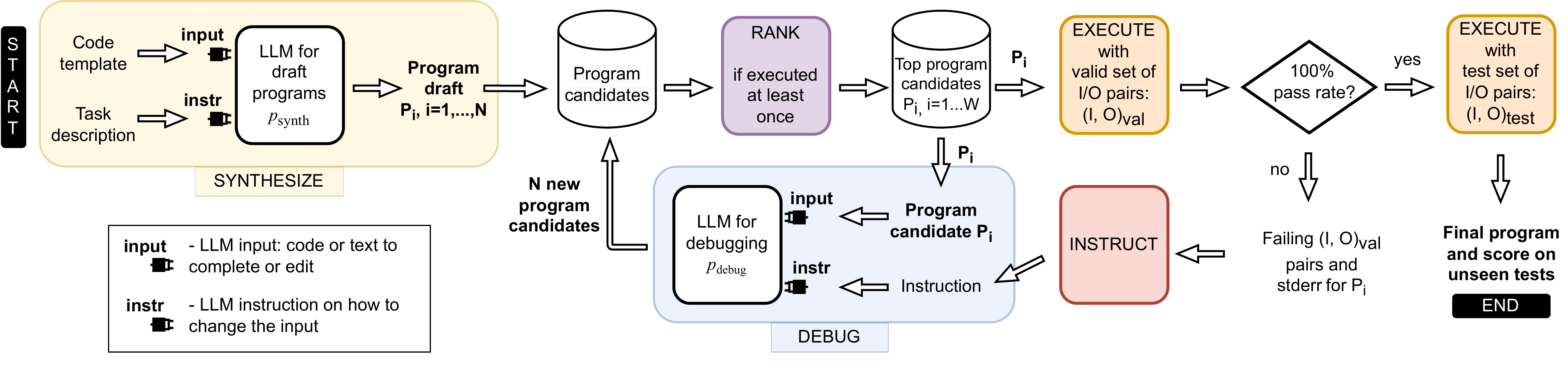}
    \caption{Framework for LLM-based Synthesize, Execute, Instruct, Debug, and Rank approach.}
    \label{fig:method}
\end{figure*}

One is \emph{Synthesize, Execute, Debug}~\cite{gupta2020:synthesize}, a framework that attempts to bridge the "last mile" gap by introducing program repair into the program synthesis algorithm. 
A programming task is specified using both a natural language description and a set of input/output (I/O) pairs demonstrating what output is expected of the program, thereby combining text to code ~\cite{iyer2018:mapping} and programming by example~\cite{halbert1984:programming,gulwani2016:programming} paradigms typical for competitive programming~\cite{zavershynskyi2018:naps}.
\emph{Synthesize, Execute, Debug} creates a first draft program using a generative model, compiles and executes it with given input examples.
This is followed by a program repair step to fix the identified errors.

Another relevant innovation is instruction-driven large language models~\cite{ouyang2022:training}. Instruction-driven models use human feedback in their training process and admit two inputs: a source text (or code) and a textual command instructing the model to edit the source in a particular way, i.e., "summarize" or "translate to Python".
These models have been shown to be highly successful in automatic program repair~\cite{fan2023:automated}. 
However, given the free-form nature of these instructions\footnote{~\ag{Throughout this paper we avoid other definitions of \emph{instruction}, such as \emph{an individual operation in code}, to prevent ambiguity.}} how one should engineer instructions that maximize repair performance is an open question. 

Section~\ref{sec:methodology} presents a framework that adapts \emph{Synthesize, Execute, Debug} to instruction-driven Large Language Models for solving programming tasks in an autonomous fashion. 
We discuss related work in Section~\ref{sec:related-work}, introduce experiments to establish optimal search and prompting strategies for this framework in Section~\ref{sec:eval}. 
Finally, we demonstrate in Section~\ref{sec:results} that our framework outperforms conventional automatic programming techniques, such as genetic programming and naive application of large language models \ag{that generate one solution per problem without updating it iteratively}. 

\section{Methodology}
\label{sec:methodology}
The proposed framework, \emph{Synthesize, Execute, Instruct, Debug and Rank}, or \method{},\footnote{~seiðr also refers to a type of Norse magic~\cite{blain2002:nine} pertaining to predicting and controlling the future, which we deem thematically appropriate.} is summarized in figure \ref{fig:method}.
To solve a programming task defined as a text description and a collection of I/O examples, we split I/O examples into prompt and validation sets and use the prompt set in a large language model to \synthesize{} a population of candidate solutions.
We \execute{} the solutions, test them against the validation set, generate a text description of the identified problems used to \instruct{} a large language model to produce repaired candidate solutions similar to the way a human developer \debug{}s a program.
We \rank{} the candidates
by correctness measured by matching I/O pairs, discard the worst candidates, and repeat until a fully correct solution is found.

\subsection{Ingredients}

\method{} makes use of 2 instruction-driven large language models for source code: a \emph{synthesis} model \synthmodel{} and a \emph{debugging} model \debugmodel{}, as well as, optionally, a large natural language model \textmodel{} that can be used for writing instructions for the code model.
Each model is a highly parameterised probability distribution over the space of (input, instruction)-tuples with parameters estimated on a large diverse (i.e., non-task-specific) corpus.
This stochastic nature of language models is an important prerequisite for \method{}, since it lets us sample batches of diverse candidate solutions from \synthmodel{}, \debugmodel{}, and \textmodel{}.
\ag{We have chosen the state-of-the-art transformer models~\cite{vaswani2017:attention} for \synthmodel{}, \debugmodel{}, and \textmodel{} in our experiments as described in Section~\ref{sec:implementation}. In general, \method{} requires a sequence-to-sequence generative model for these blocks.}

\subsection{Synthesize}
\label{sec:synth}

\begin{figure}[b]
    \centering
    \includegraphics[width=\linewidth, trim={0mm 33mm 0mm 6mm}]{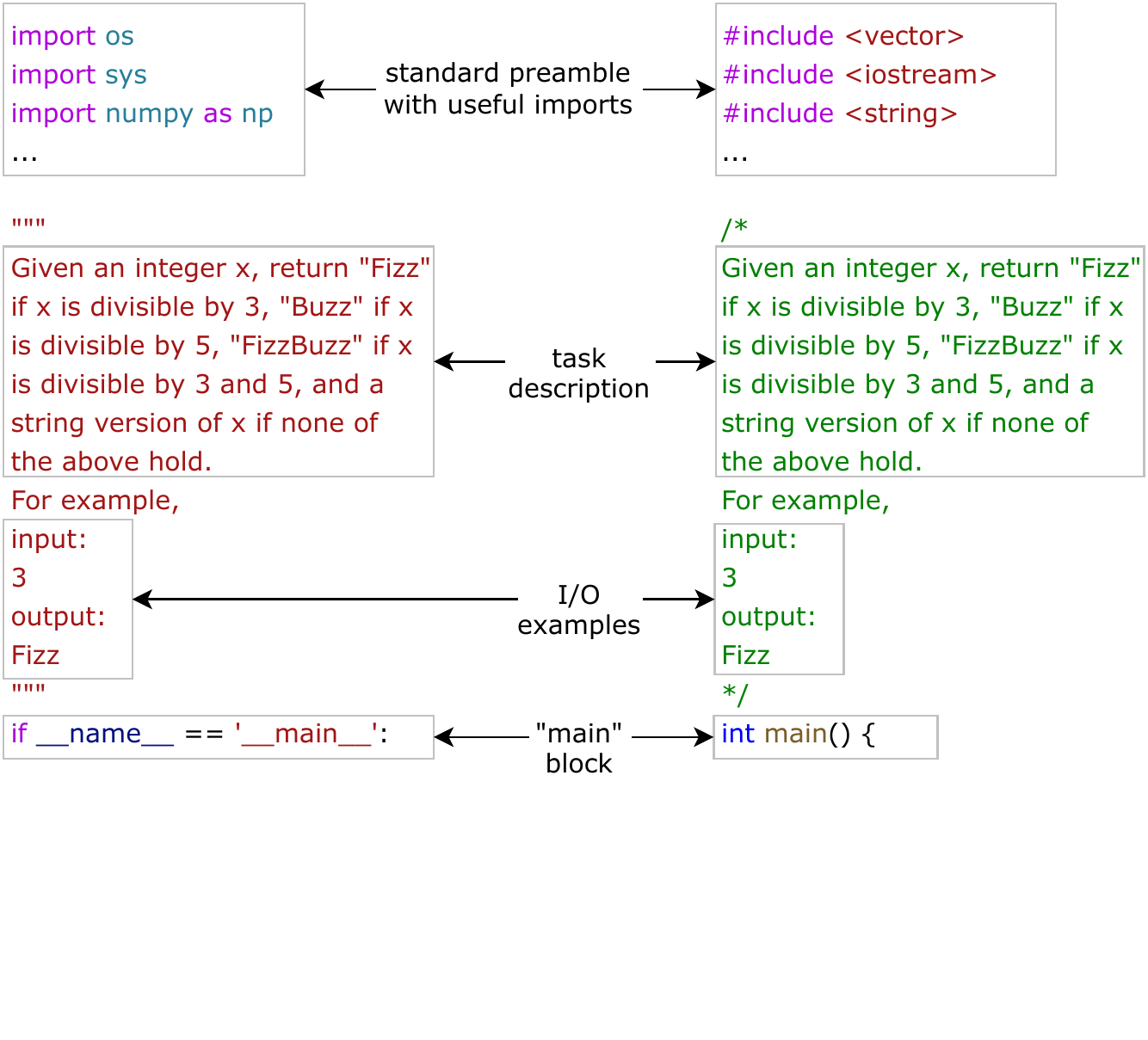}
    \caption{Anatomy of \synthesize{} templates}
    \label{fig:template}
\end{figure}

\begin{figure*}[t]
    \centering
    \includegraphics[width=\textwidth,trim={0mm 6mm 0mm 0mm}]{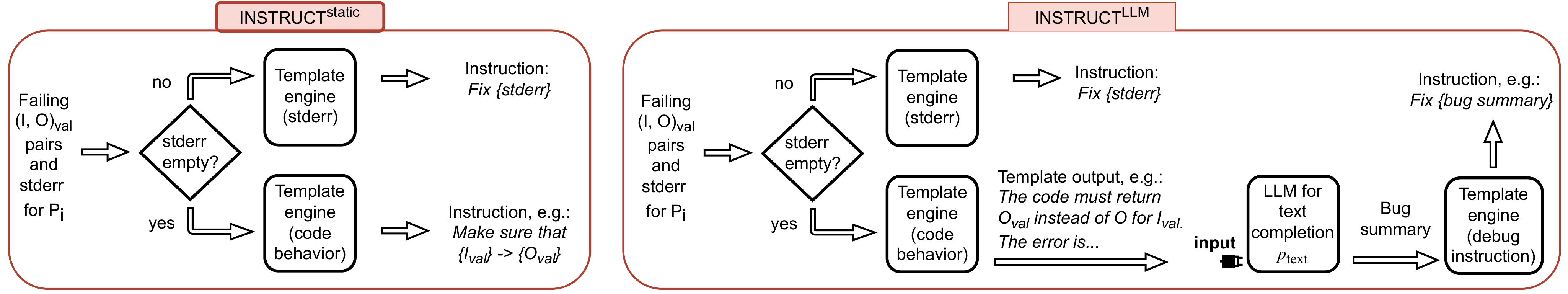}
    \caption{\instruct{} blocks with and without an LLM.}
    \label{fig:method-instruct}
    \vspace*{-3ex}
\end{figure*}

The framework starts with the \synthesize{} block, which is responsible for generating initial draft solutions to programming tasks to be repaired in the later stages of \method{}.
We start with a basic template for a chosen programming language that contains a number of standard library imports and an empty \emph{main} function or this language's equivalent thereof, see figure~\ref{fig:template}.
We populate this template with a comment indicating a text description of a task at hand and several I/O examples from the prompt training set.
We design the templates with guidelines by the authors of the language model~\cite{:openai} and prior work~\cite{debruin2021:autoencoders} in mind.
We then sample $\treearity{}$ programs from \synthmodel{}, setting \texttt{input} to the populated template and \texttt{instruction} to the problem description.
We use temperature sampling with a monotonically increasing temperature schedule where $i$-th program is sampled with temperature $t_i \approx \frac{i}{N}$ (approximate equality enables efficient implementation by means of batching).
Thus, the sampling procedure for the first programs approximates deterministic maximum likelihood estimation.
Ultimately, this approach ensures that samples are diverse, but always contain the likeliest programs.

\subsection{Execute}

In the \execute{} block, the programs are compiled (if necessary) and launched using the standard tools for the programming language.
The program is run once for every I/O pair in the validation set. 
Its \texttt{stdin} stream receives all the input lines in a given input pair, and its \texttt{stdout} and \texttt{stderr} streams are captured and saved.
We then measure the \emph{score} of the program defined as accuracy over output lines, with \expectedoutput{} being the expected output, and $n=\max\{|\expectedoutput{}|, |\text{stdout}|\}$:
\[    
\text{score}(\expectedoutput{}, \text{stdout}) = \frac{\sum^{n}_i{\mathbb{I}[\text{stdout}_i = O_i]}}{n} 
\]
unless \texttt{stderr} is non-empty during compilation or execution, which is considered to indicate failure and is assigned a score of 0.

\subsection{Instruct}
\label{sec:instruct}

The goal of the \instruct{} block is to provide an instruction that summarizes a bug in the program candidate for \debugmodel{}. 
The resulting instruction with the bug summary should indicate what requirement is violated and instruct the LLM to edit the candidate program so that the candidate meets the violated requirements. 
In \method{}, we generate instructions using template engines. 
In general, template engines replace placeholders in files or strings with input values and return a formatted string. 
With template engines, we can create a number of templates that will be adapted dynamically based on the results of program candidate execution. 

We consider two different designs of the instruction generation block: \instructs{} and \instructllm{} shown in figure~\ref{fig:method-instruct}. 
Both types of blocks use failing I/O pairs from the validation set and \texttt{stderr} output of the candidate execution. 
In both blocks, if \texttt{stderr} is not empty, i.e., execution errors occur before getting the output to compare it with the expected output, the \texttt{stderr}-based template engine generates an instruction to fix the error mentioned in \texttt{stderr}. 
However, the blocks differ in the way they transform failing I/O pairs to generate instructions in case \texttt{stderr} is empty.

\instructs{} uses a fixed input template and substitutes placeholders for input and output with the corresponding strings of the first failing test case.
We show the resulting instruction for an exemplar template in figure~\ref{fig:method-instruct}.
By contrast, \instructllm{} uses the failing I/O pair in the LLM for text completion, thereby prompting the text LLM to produce the bug summary. 
An exemplar output of the code behavior template engine in figure~\ref{fig:method-instruct} describes that the code returns output O instead of expected output O$_{\text{val}}$ for the failing test case with input string I$_{\text{val}}.$
The LLM is then prompted to auto-complete this description of program behavior with the bug summary. 
The bug description is passed further to the next template engine and used as the debugging instruction, such as ``\emph{Fix~\{bug summary\}}''.

\subsection{Debug}

The \debug{} block iterates over all programs in the population and uses the instruction written by \instruct{} based on the results of \execute{} to sample from \debugmodel{} $\treearity{}$ times
to repair every candidate, setting \texttt{input} to the candidate solution and \texttt{instruction} to the output of \instruct{}.
The population of candidates is then replaced with the output of \debug{}.

\subsection{Rank}

Finally, the \rank{} block implements what is known in genetic programming as \emph{\ag{parent} selection}~\cite{koza1994:genetic}.
It ranks all programs in the candidate population by their score calculated in \execute{}, keeps the top $\beamwidth{}$ programs, and removes the rest from the population.

\subsection{Meaning of Hyperparameters}
\label{sec:beam-search}

\begin{figure}
    \centering
    \includegraphics[width=\linewidth, trim={0mm 4mm 0mm 0mm}]{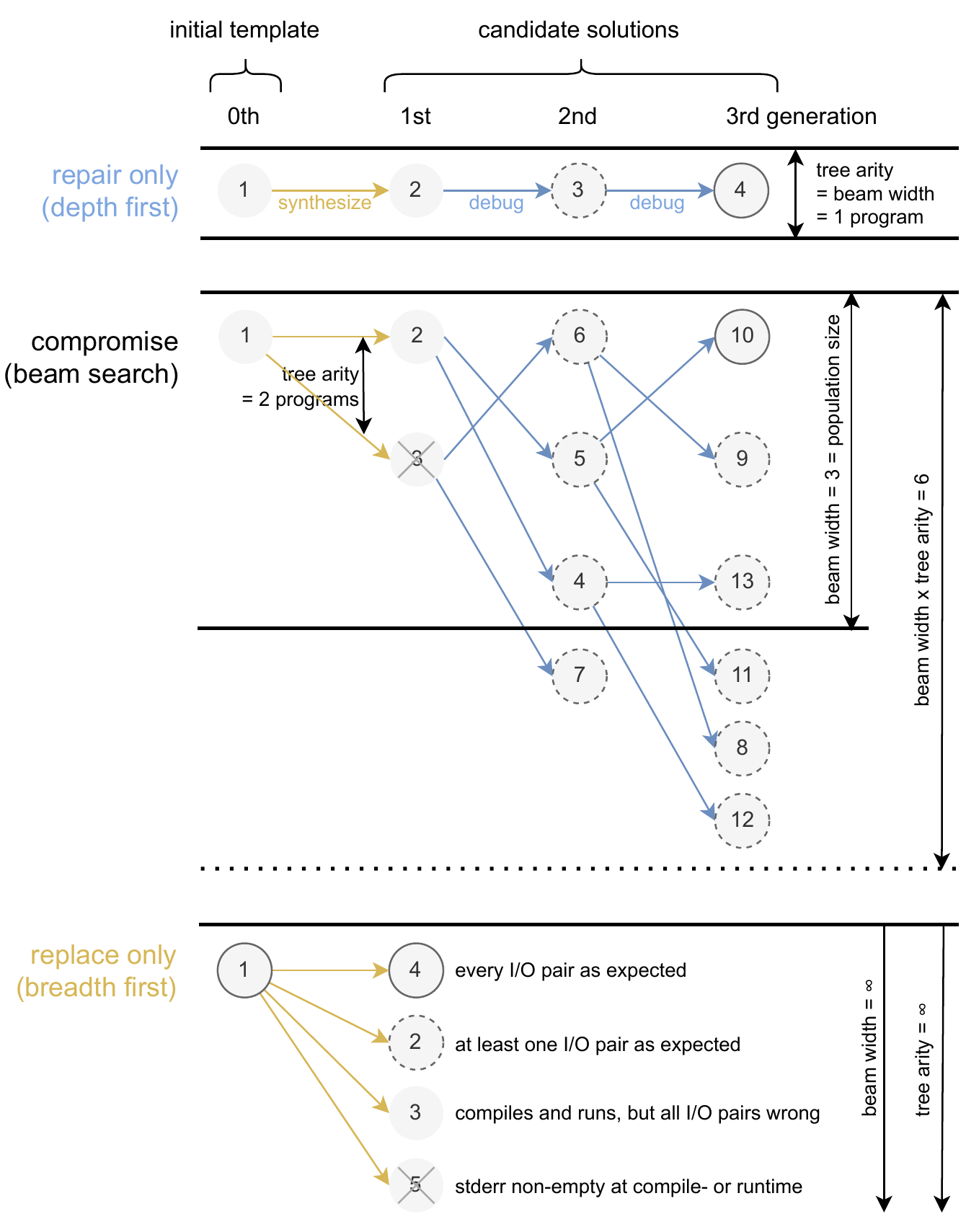}
    \caption{Repair-replace trade-off as a tree search problem.}
    \label{fig:beam-search}
    \vspace*{-4ex}
\end{figure}

After evaluating a given candidate solution in \execute{}, \method{} supports two approaches to addressing the candidate's flaws:
\begin{itemize}
  \item \emph{Replace} the candidate with another sample from the current population.
  \item Use \instruct{} and \debug{} to repair the candidate.
\end{itemize}
We refer to this problem as \emph{repair-replace trade-off}, by analogy with production economics~\cite{jack2000:optimal}. 

How does the choice of hyperparameters $\treearity{}$ and $\beamwidth{}$ influence the flow of \method{}?
$\treearity{}$ and $\beamwidth{}$ act as upper bounds on the \emph{replace} option by limiting the size of the population.
In the edge cases, $\treearity{} = \beamwidth{} = 1$ corresponds to a repair-only process, while $\treearity{} = \beamwidth{} = \infty$ corresponds to replace-only, see figure~\ref{fig:beam-search}. 

Observe that a mutation-only genetic algorithm with population size $\beamwidth{}$ , such as \method{}, is equivalent to \emph{local beam search} with beam width $\beamwidth{}$ on a $\treearity{}$-ary tree~\cite[Section 4.1.4]{russell2010:artificial}. This corresponds to a known property of local beam search: it degenerates into depth-first search when $\beamwidth{} = 1$, whereas setting $\beamwidth{} = \infty$ yields breadth-first search.
Hence, we refer to $\treearity{}$ as \emph{tree arity} and $\beamwidth{}$ as \emph{beam width}.

\section{Related Work}
\label{sec:related-work}

The use of large language models for a program repair step within a program synthesis pipeline has been studied by \citex{joshi2022:repair} and \citex{gupta2020:synthesize}, 
while a specific case of instruction-driven LLMs has been explored by~\citex{fan2023:automated}.
The latter authors also compare instruction-driven LLMs to other Automated Program Repair (APR) strategies, and conclude that \ag{LLMs solve the task effectively.}
However, they do not consider the entire \emph{Synthesize, Execute, Debug} framework, as we do in this work. 
\ag{By contrast, LLMs are usually set up to generate a fix from the first debug attempt and do not update failing patches iteratively.}

Prompt generation for LLMs pre-trained on code is investigated in prompt engineering literature on a number of tasks, such as source code repair, type inference, code synthesis, and autocompletion~\cite{shrivastava2022:repositorylevel, huang2022:prompttuned, ahmad2023:fixing}.
Studies on program repair automation used prompts that contain code only, code with docstrings, and code with bug hints with the Codex model to test the repair capabilities of the LLM on the QuixBugs benchmark~\cite{lin2017:quixbugs, prenner2022:can}. 
The studies reported that bug localization hints were not helpful, whereas providing buggy code and the task summary was the most effective. 
ChatGPT was tested on QuixBugs in addition to Codex models as well~\cite{sobania2023:analysis}.
Kuznia et al.~\cite{kuznia2022:less} proposed to summarize task descriptions of competitive programming and interview programming tasks. 
Following guidance from these studies, we include I/O pairs and task descriptions as docstrings in addition to the function signature in our prompts. 

\section{Expermental Setup}
\label{sec:eval}

To explore the capabilities of \method{}, we test the framework on the benchmark of problems for code competitions with different types of instructions for program candidate debugging, varied search strategies, and two languages, Python and C++. 
Our experiments use the Program Synthesis Benchmark~2 (PSB2) for problem descriptions and tests to evaluate the proposed framework~\cite{helmuth2022:applying}. 
We compare the performance of programs synthesized with \method{} to the PushGP genetic programming system with down-sampled lexicase selection~\cite{helmuth2022:problemsolving}. 
During our empirical evaluation of \method{} performance, we address the following research questions:
\head{RQ1. Repair-replace trade-off exploration} 
What is the impact of using different tree search strategies
in the autonomous programming setting? 
We experiment with four different tree arities in the tree search and study their impact on the number of resolved problems as well as the speed of obtaining solutions.
\head{RQ2. Prompt engineering} What is the effect of using LLM-produced bug summaries compared to static instructions on the repair performance of automatically synthesized code? We test six static debug instructions that describe bug behavior based on violated requirements and five auto-generated debug prompts. 

\subsection{Data}
\label{sec:data}

PSB2 is a benchmark suite of 25 problems for program synthesis that resemble small real-world tasks. PSB2 was developed as a more realistic and challenging version of PSB1~\cite{helmuth2015:general}, the latter consisting of textbook problems and is widely used in genetic programming~\cite{sobania2022:choose}. 
The problems require different data structures and control flows to be used for effective solutions and are taken from sources, such as competitive programming platforms and educational courses. 
The problems have descriptions in English, as well as 1 million~(M) tests for training and 1M testing-stage tests, including edge or corner cases that test the resulting program on complicated inputs. 
The tests are provided as I/O pairs and are distributed together with the problem descriptions as a PyPI package.\footnote{~\url{https://pypi.org/project/psb2/}} 

In PSB1~\cite{helmuth2015:general}, the training set consists of the edge test cases and is augmented by random test cases if the number of edge tests is not enough. The test set is formed by random test cases. 
This terminology is preserved in PSB2. 
However, we do not have a training or fine-tuning phase in our experiments, \ag{because the models are not made available for further training}. Instead, we validate the framework with an existing pre-trained LLM for code and text as its parts. 
Therefore, we only have the validation and test phases. 
We will refer to training test cases in the PSB terminology as validation test cases in this study. 

\subsection{Repair-replace Trade-off Settings}
\label{sec:trade-off-settings}

As described in Section~\ref{sec:beam-search}, the choice of beam width $\beamwidth{}$ and tree arity $\treearity{}$ define the repair-replace trade-off where higher $\beamwidth{}$ and $\treearity{}$ prioritize to repair over replace. 
We evaluate four options for these hyperparameters as shown in table~\ref{tab:w-n}. 

\begin{table}[h]
    \centering
    \vspace*{-1ex}
    \caption{Tree search hyperparameters.}\small
    \label{tab:w-n}\vspace*{-4mm}
    \begin{tabular}{r|c|c|c|c}
    experiment & 1 & 2 & 3 & 4 \\
    \midrule
     beam width $\beamwidth{}$ & 1 & 10 & 100 & $\infty$ \\
     tree arity $\treearity{}$ & 1 & 10 & 100 & $\infty$
    \end{tabular}
    \vspace*{-1.8ex}
\end{table}

Because we aim to compare tree search parameters, we fix one default debugging instruction and use the \instructs{} block. 
Moreover, we set the upper limit for the total number of generated program candidates to 1000 to limit the experimentation time. 
Although some solutions may not be found within the hard limit, we assume\footnote{~This assumption is later confirmed in Section~\ref{sec:rq1}.} that 1000 program candidates form a sufficiently large search space for our experiments.
$\beamwidth{} = \treearity{} = \infty$ is achieved in implementation by setting $\beamwidth{}$ and $\treearity{}$ equal to the upper limit on the number of candidates, i. e. 1000.
This setting ensures that a second generation of programs does not exist.

\subsection{Prompting Strategies}
The prompt for the LLM model \debugmodel{} consists of the input for editing --- candidate program generated so far --- and a debug instruction to repair the candidate. 
We test \method{} on 11 debug instructions to explore whether the use of the LLM for text completion \textmodel{} benefits the performance of our framework, as well as what effect different prompt phrases have on the debug process. 
We compare debug instructions that use neutral phrases with those that use more confident language and mimic experienced software developers, as well as shorter and longer instructions with different amounts of details about code behavior.
To alleviate the effect of beam width and tree arity, we set $N=W=1$ and test the repair-only tree search strategy shown in figure~\ref{fig:beam-search}. 
This strategy is used to gradually improve one program candidate throughout the search with no competing programs in the same generation. 

The debug instructions are formulated as templates. The instructions describe the violated requirements in terms of the wrong output in a failing I/O test or summarize the bug to capture issues in code logic.
We present debug instructions using the template engine format: the brackets \{ \} denote that the placeholder in the brackets will be replaced with the value generated during execution, \{I$_{\text{val}}$\} and \{O$_{\text{val}}$\} stand for values failing I/O pair from the validation set.
As shown in figure~\ref{fig:method-instruct}, the instruction to fix the execution errors, which abort the program before the resulting output is obtained, with \texttt{stderr} lines: Fix \{stderr\}. 
Static debug instructions that do not use LLM for bug summarization are as follows:
\begin{enumerate}[label=S\arabic*]
\setcounter{enumi}{-1}
    \item \label{prompt-0} Make sure that \{I$_{\text{val}}$\} -> \{O$_{\text{val}}$\};
    \item \label{prompt-1} Make sure the code returns \{O$_{\text{val}}$\} for input \{I$_{\text{val}}$\};
    \item \label{prompt-2} Ensure that input \{I$_{\text{val}}$\} yields output \{O$_{\text{val}}$\};
    \item \label{prompt-3} Modify code to get \{O$_{\text{val}}$\} from \{I$_{\text{val}}$\};
    \item \label{prompt-4} Code must correspond instructions in comments and \{I$_{\text{val}}$\} must yield \{O$_{\text{val}}$\};
    \item \label{prompt-5} See comments in code and return \{O$_{\text{val}}$\} for input \{I$_{\text{val}}$\}.
\end{enumerate}

The instruction \ref{prompt-0} is the default instruction for tree arity experiments. 
It has an intuitive symbolic notation (->) instead of the word ``return'' or ``yield''. 
In instructions \ref{prompt-1}--\ref{prompt-3}, we experiment with verbs and the order of output and input. 
Alternatively, in debug instructions \ref{prompt-4}--\ref{prompt-5}, we prompt the model to consider task description in the docstring in addition to providing the details of the failing I/O pair. 
Overall, instructions \ref{prompt-0}--\ref{prompt-5} indicate the requirements to be met, but do not describe the current program's behavior. 

The second set of instructions use the LLM for text completion \textmodel{}. 
The instructions are designed so that the LLM is prompted to complete the sentence that should describe an error. 
In addition to validation I/O pairs, the following notation is used: \{O$_{\text{p}}$\} denotes the program candidate output for input \{I$_{\text{val}}$\}, \{task\} is a placeholder for a problem description in English. 
\ag{Note that we do not include the incorrect output $O_p$ of a generated candidate program in debug instructions S0-S5, because it is recommended to avoid asking the model what not to do.\footnote{~\url{https://help.openai.com/en/articles/6654000-best-practices-for-prompt-engineering-with-openai-api}}}
We denote the text completion LLM's output as \{bug\} which should constitute the bug summary.
Input templates to use LLM for bug description followed by debugging instruction templates (after ``$\rightarrow$'') are as follows:
\begin{enumerate}[label=M\arabic*]
\setcounter{enumi}{5}
    \item \label{prompt-6} The code should solve the following problem: \{task\}. The code must return \{O$_{\text{val}}$\} for input \{I$_{\text{val}}$\} but it returns \{O$_{\text{p}}$\}. Obviously, the error is that... \newline 
    $\rightarrow$ Fix \{bug\};
    \item \label{prompt-7} The code should solve the following problem: \{task\}. The code must return \{O$_{\text{val}}$\} for input \{I$_{\text{val}}$\} but it returns \{O$_{\text{p}}$\}. The error is that... \newline 
    $\rightarrow$ Fix \{bug\};
    \item \label{prompt-8} Problem description: \{task\}. The code must return \{O$_{\text{val}}$\} for input \{I$_{\text{val}}$\}, but it returns \{O$_{\text{p}}$\}. It is clear the error is that... \newline 
    $\rightarrow$ Fix \{bug\};
    \item \label{prompt-9} There is clearly a bug in code, because the code returns \{O$_{\text{p}}$\} for input \{I$_{\text{val}}$\} but output \{O$_{\text{val}}$\} is expected. The bug is that... \newline 
    $\rightarrow$ Fix \{bug\};
    \item \label{prompt-10} There is clearly a bug in code, because the code returns \{O$_{\text{p}}$\} for input \{I$_{\text{val}}$\} but output \{O$_{\text{val}}$\} is expected. The bug is that... \newline 
    $\rightarrow$ Fix \{bug\} and modify the code to return \{O$_{\text{val}}$\} for input~\{I$_{\text{val}}$\}.
\end{enumerate}
Note that the text completion LLM does not use program candidates in its input, but only template inputs \ref{prompt-6}--\ref{prompt-10} before the arrow. 

Input \ref{prompt-6} for the text completion LLM is used to evaluate the effect of the ``confidence'' sentiment on the bug summaries and debugging process. 
It is identical to input \ref{prompt-7}, except for the word ``obviously'', which should reflect or confidence of the comment. 
Inputs \ref{prompt-7} and \ref{prompt-8} can be compared in the way the problem description is introduced, i.e., as a separate sentence similar to a spoken situation in prompt \ref{prompt-7} or as a short title in \ref{prompt-8}.

Input templates \ref{prompt-9} and \ref{prompt-10} for text completion LLM are identical, but the instruction templates are different.
Text completion inputs start with a ``confidently'' phrased statement that a bug is present in code.  
We include both the LLM output \{bug\} and description of the failing validation test case in debug instruction \ref{prompt-10}.
Therefore, instructions \ref{prompt-6}--\ref{prompt-9} rely mainly on the LLM output to summarize the bug, whereas instruction \ref{prompt-10} also provides information about the expected output. 

\subsection{Performance Indicators}
\label{sec:metrics}

\sloppy %
In our experiments, we compare 
the number of fully solved programs obtained using \method{} with different values of hyper-parameters. 
For a more detailed analysis of results, we use \emph{test pass rate (TPR)} and \emph{Excess Programs Generated (EPG)}.
TPR reflects the percentage of fully passed test cases based on the exact match of program output and test output. 
The TPR metric is used for the final evaluation of generated programs and does not reflect partial passing of the I/O test as opposed to \emph{score} in the \rank{} block. 

\debug{} and \execute{} blocks generate a number of programs that are replaced or repaired during the search for solution program. 
The number of programs generated before the first occurrence of the program that passes all validation test cases is referred to as EPG. 
EPG is indicative of the computational cost of solving a problem distributed in terms of LLM inferences and program compilations and executions.

\subsection{Implementation Details}
\label{sec:implementation}

We use GPT-3 models pre-trained on code\footnote{~\url{https://platform.openai.com/docs/guides/code/editing-code}} and text\footnote{~\url{https://platform.openai.com/docs/models/gpt-3}} as LLMs in our framework. 
Specifically, we use Codex-edit (code-davinci-edit-001) as the LLM for draft programs $p_\text{synth}$ and LLM for debugging $p_\text{debug}$ and GPT-3 (text-davinci-003) for bug summarization via text completion with $p_\text{text}$. 
We ensure that the program candidates generated from the same parent program are different from each other by changing the temperature parameter of Codex-edit. 

We use 2000 I/O pairs from the test split of PSB2 to evaluate the candidate program that has passed all the validation test cases during debugging. 
Due to repetitive calls to the \execute{} block, we have to resolve the speed of testing versus precision trade-off while choosing the number of validation test pairs.
We resolve the trade-off by fixing the validation set size at~100.

In the experiments with tree arity values, we set the limit to generate a maximum of 1000 program candidates during the search of the candidate that passes all validation tests. 
If we reach 1000 candidates and none of them passes all validation tests, we report the test pass rate for the last generated candidate. 
In the experiments with prompts, we set the limit of maximum generated programs to~5, because we search for the prompt that yields the fastest solution to exclude long searches \ag{and comply with the request rate limits}.

\section{Results and Discussion}
\label{sec:results}

\subsection{RQ1. Repair-replace Trade-off Exploration}
\label{sec:rq1}
\begin{figure}[t]
  \centering
  \includegraphics[width=0.8\linewidth, trim={0mm 2.8mm 0mm 2mm}, clip]{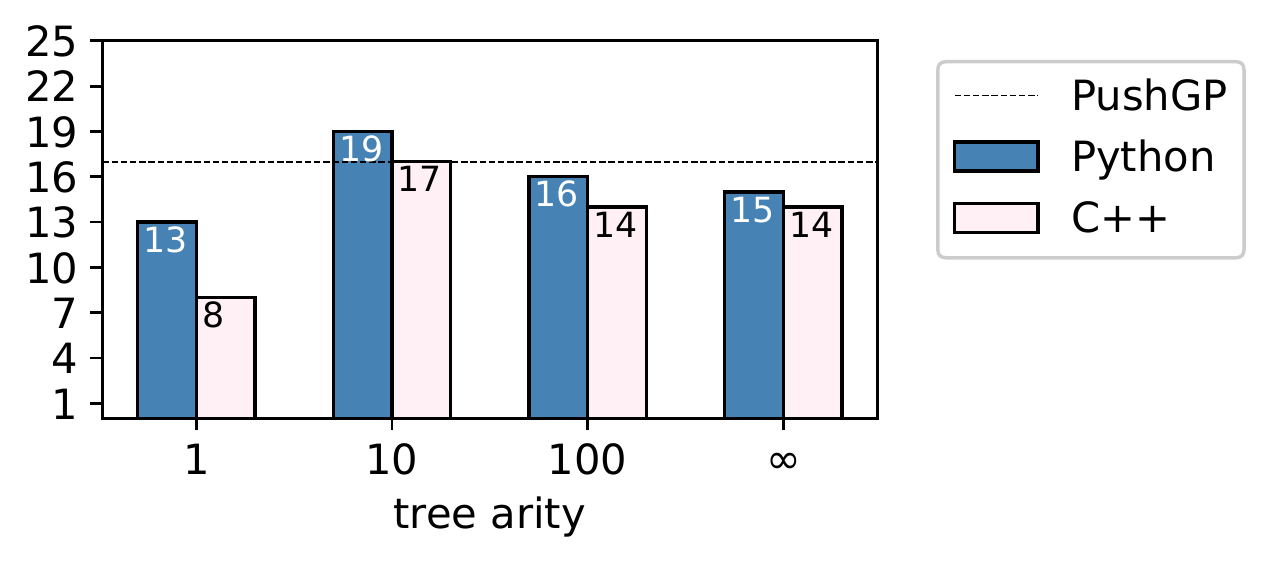}
  \vspace*{-2mm}
  \caption{Number of solved PSB2 problems depending on the tree arity in tree search for the fixed prompt type \ref{prompt-0}.}
  \label{fig:solved-vs-bf}
  \vspace*{-2ex}
\end{figure}

We compare the number of solved problems in the experiments with tree arity of 1, 10, 100, and $\infty$ and fixed debug instruction \ref{prompt-0} in Python and C++ in figure~\ref{fig:solved-vs-bf}. 
The results of \method{} are compared to the baseline performance of PushGP on the PSB2 benchmark, which solves 17 out of 25 problems. 
Note that experiments with $N=1$ and $N=\infty$ can be considered as ablation studies, where the replace option and repair option is turned off, correspondingly. 

The results highlight the benefit of compromise strategies with tree arity of 10 and 100 over repair-only ($N=1$) and replace-only ($N=\infty$) strategies. 
The repair-only scheme is outperformed by other strategies. 
We explain the poor performance of repair-only strategy by the fact that the search space is under-explored. 
Specifically, replace scenario ensures the LLM for debugging represented by Codex-edit in our experiments generates different updates of program candidates using variable temperature.
The probability of finding a better fix is higher when more alternatives are generated to update the draft program at $N>1$ compared to $N=1$. 
The search strategy with $N=10$ yields the best results: it performs on par with PushGP for C++ and outperforms the baseline during Python program synthesis by +2 problems resulting in a total of 19 programs that pass all test cases.
The results imply that generating a moderate number of programs in parallel during the \debug{} step works better than the policies in which more updates are generated for each program (100 or 1000) or only one program is updated iteratively.

\begin{figure}[t]
\begin{subfigure}[t]{\columnwidth}
\centering
\includegraphics[width=\linewidth, trim={0mm 4mm 0mm 0mm}]{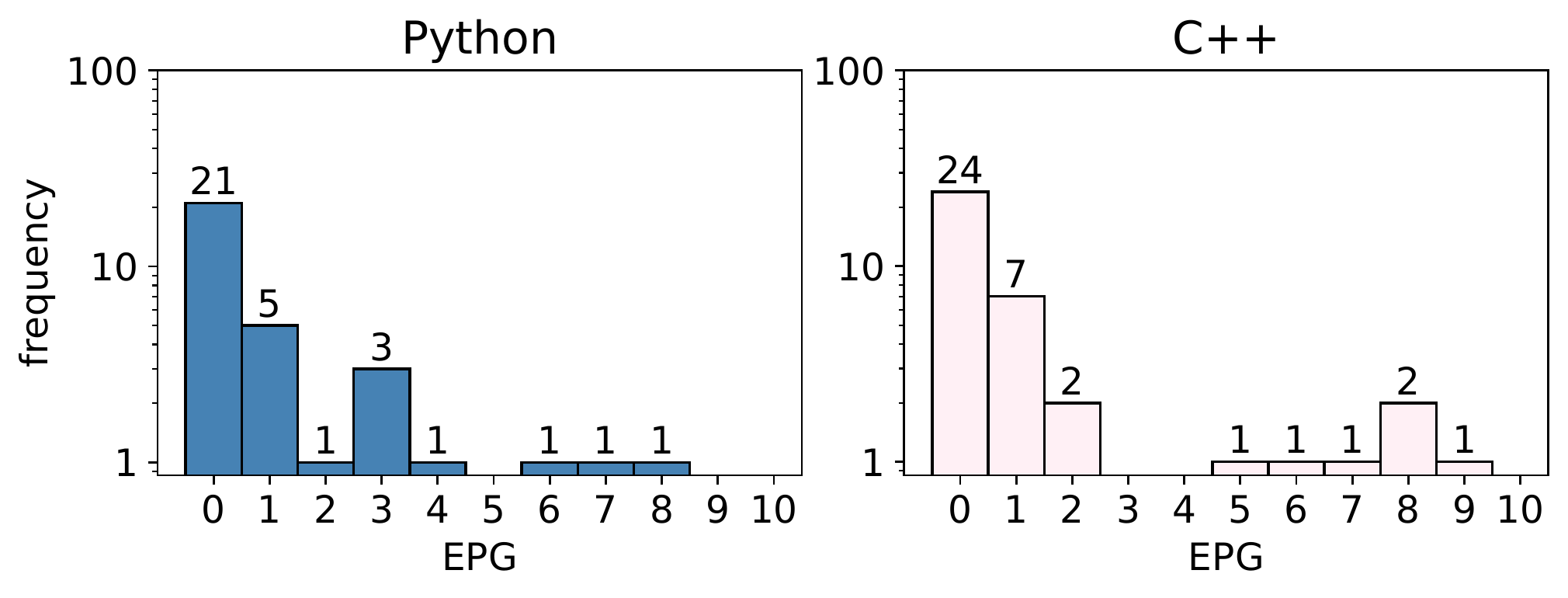}
  \caption{0 $\leq$ EPG $\leq$ 10 with step 1.}
  \label{fig:epg-distrib-solved-10}
\end{subfigure}

\vspace{2mm}

\begin{subfigure}[t]{\columnwidth}
\centering
\includegraphics[width=\linewidth, trim={0mm 4mm 0mm 0mm}]{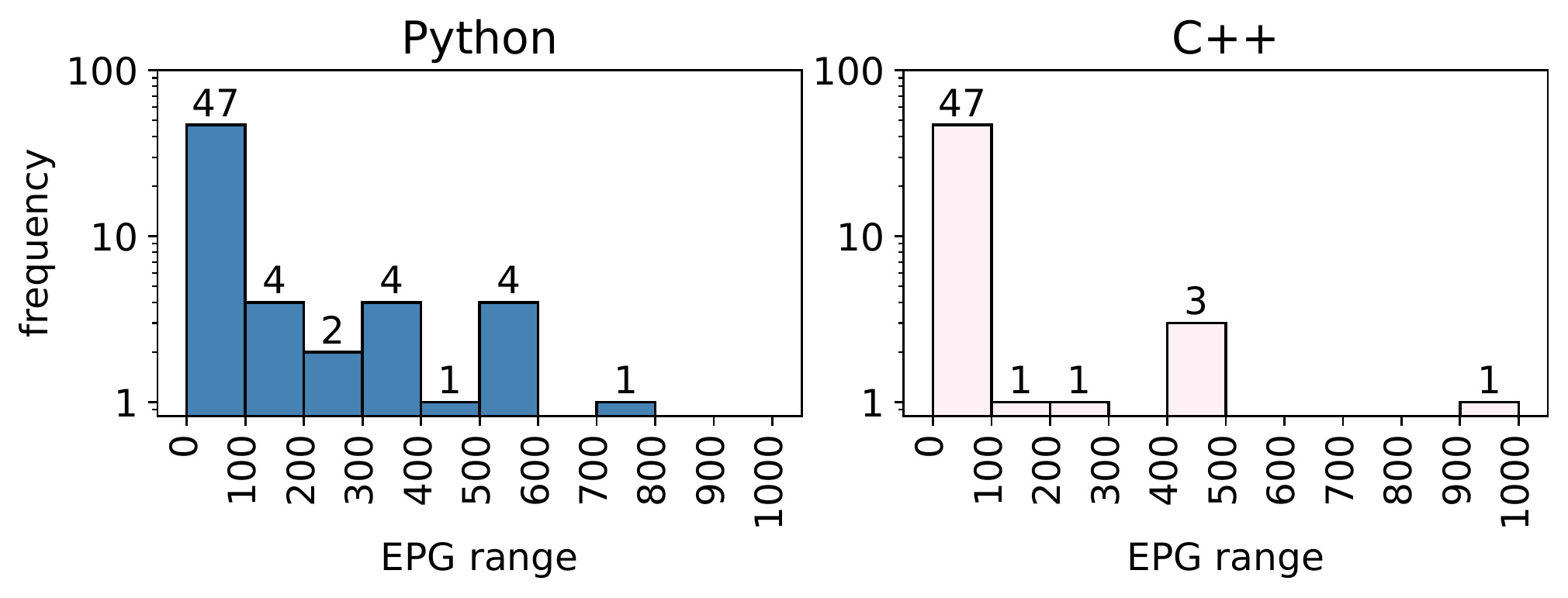}
  \caption{0 $\leq$ EPG $\leq$ 1000 with step 100.}
  \label{fig:epg-distrib-solved-100}
\end{subfigure}
\vspace{-2mm}
\caption{Distribution of the number of generated programs during each problem-solving attempt in the experiments with different tree arities where a problem solution is found.}
\label{fig:epg-distribution}
\vspace{-2ex}
\end{figure}

We present the analogy of the solution speed for all four arities and fixed default debug instruction in figure~\ref{fig:epg-distribution}. 
In detail, we show the distribution of EPG values in all experiments to explore how many candidate updates are generated before the solution is found.
We zoom in to the cases with solutions found with up to the first 10 program candidates in figure~\ref{fig:epg-distrib-solved-10} and show the EPG distribution with the step of 100 candidates in figure~\ref{fig:epg-distrib-solved-100}. 

Out of 100 experiments for each language, in 21--24\% of runs in Python and C++, the draft program is already the solution (EPG=0). 
For 19-32\% of experiments, the solution is found after discarding 5 candidates. 
Around half of experiments do not generate more than 100 programs. 
However, 5 problems are solved with more than 500 generated programs in Python and 1 problem in C++ (with $N=10$).
The results imply that the first steps in the update of the draft program are crucial for solving the problem. 
The chances of solving the problem on the later stages of the search, such as after 100 programs have been generated, are low.
This confirms our initial assumption in Section~\ref{sec:trade-off-settings} that 1000 programs are sufficient.

\begin{mdframed}[style=mystyle]
\noindent
\textbf{Answer to RQ1.} 
\method{} outperforms the PushGP baseline on PSB2 in Python and performs on par with it in C++ experiments with tree arity of 10. 
Search strategies with tree arity larger than one benefit from the replace possibility of the \method{} framework as a consequence of using variable temperature for Codex-edit.
The repair component is also crucial for the framework because the replace-only search policy (with tree arity of $\infty$) performs worse than the policies that alternate between replace and repair during program update (with tree arity of 10 or 100).  
\end{mdframed}

\subsection{RQ2. Prompt Engineering}
\begin{figure}[t]
  \centering
  \includegraphics[width=\linewidth, trim={0mm 2mm 0mm 2mm}, clip]{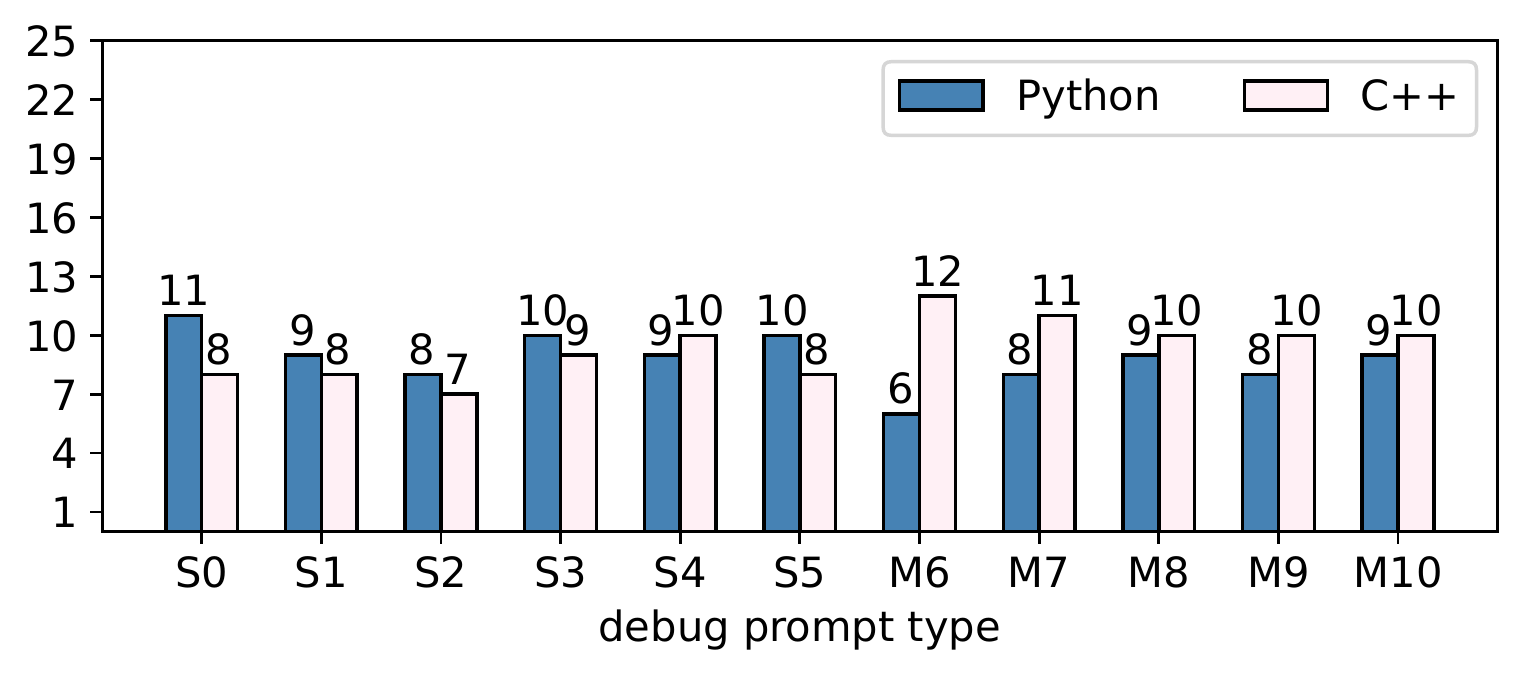}
  \vspace*{-7mm}
  \caption{Number of solved PSB2 problems depending on the instruction choice for the fixed tree arity of 1. 
  }
  \label{fig:solved-vs-prompt-id}
  \vspace*{-3.8ex}
\end{figure}

We report the number of solved problems for different static and GPT-assisted debug instructions in figure~\ref{fig:solved-vs-prompt-id}. 
Because debug instructions are parts of prompts for LLMs and the program candidate format does not change, we will use the term \textit{prompt} during the analysis of experiment results with different instructions.
Overall, the performance of the framework is robust to the debug prompt choice, both with LLM-generated and static templates. 
The number of solved problems differs for Python and C++ in our experiments.

For C++, all debug prompts except \ref{prompt-2} result in the same or higher performance than the instruction \ref{prompt-0} which is used in the repair-replace trade-off experiments. 
The debug instruction \ref{prompt-2} contains the verbs ``yield'' and ``ensure'' which are probably rarely used in code documentation. 
The best debug instruction for C++ is the LLM-assisted template \ref{prompt-6} containing the word ``obviously'', which should indicate the confidence of the author of bug summary whom GPT-3 should mimic during autocompletion.

\begin{figure*}[t]
  \centering
  \includegraphics[width=.8\textwidth, trim={0mm 2mm 0mm 2mm}, clip]{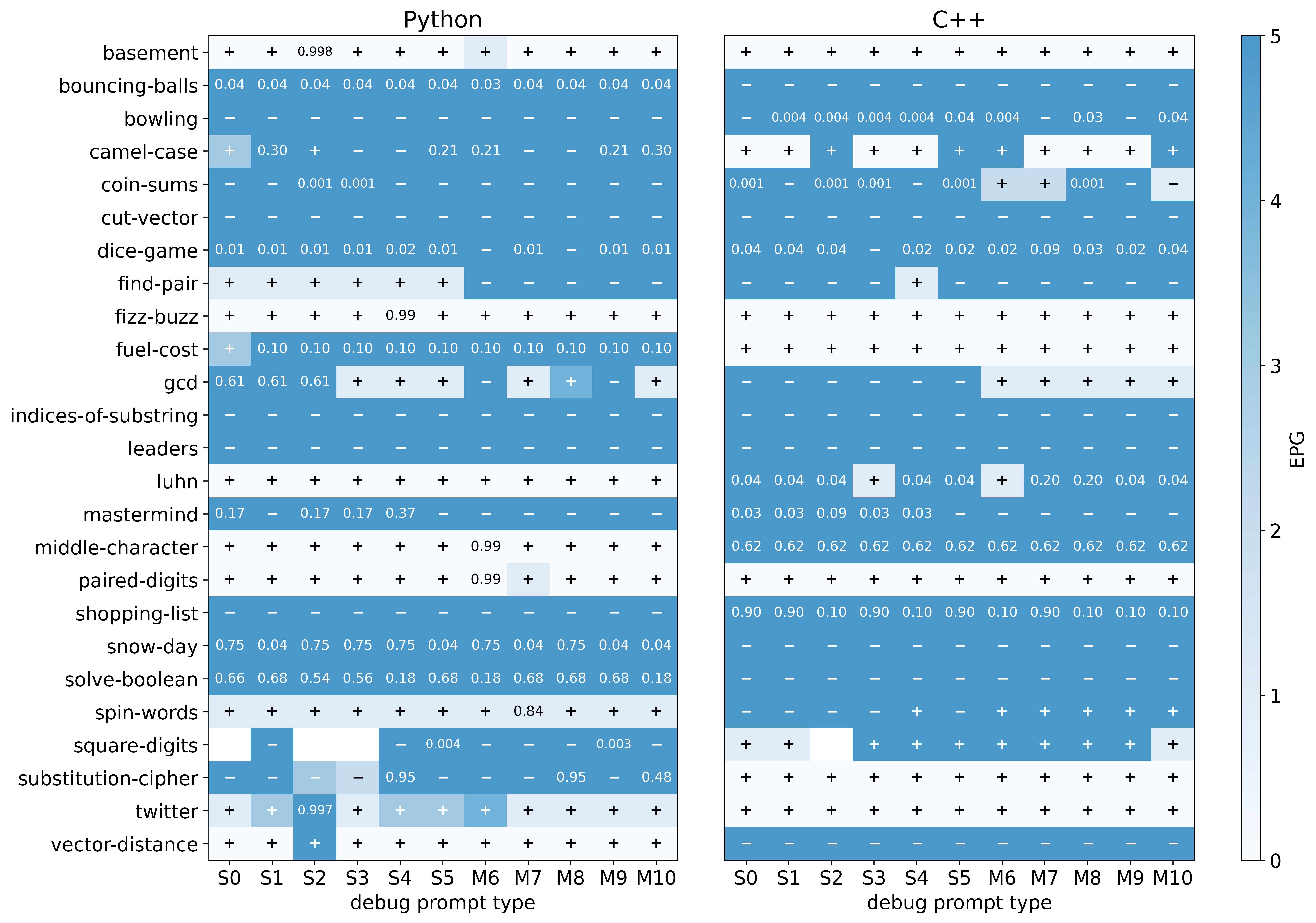}
  \vspace*{-2mm}
  \caption{Number of excess programs generated (in color) and test pass rate (as numbers) depending on the type of debug prompt. Higher EPG values are shown in darker shades than low EPG. We denote solved problems with ``+'' (test pass rate = 1), unsolved problems with ``-'' (test pass rate = 0), and show the test pass rate for partially solved problems. }
  \label{fig:epg-prompt-test}
\end{figure*}

Python programs do not show the same effect during experiments with different prompts. 
The overall performance drops in comparison with using the prompt \ref{prompt-0}. 
By limiting the total number of generated programs from 1000 
to 5 in the current set of experiments, we lose 2 problem solutions in Python with \ref{prompt-0}. 
The prompt that results in the best performance in C++ for the EPG limit of 5 corresponds to the worst performance in Python. 
This result can occur due to the small tree arity and low variability of debugging updates of the initial draft. 
Another reason is that the GPT-3 summary of bugs may not point to logical errors. The model for text autocompletion frequently outputs bug summaries that mention ``the code is not accepting the input correctly.''
Note that such bug summary appears in other debug prompts, too. 

To analyze the effect of using different prompts on a problem level, we present a heatmap of EPG for all 25 problems in figure~\ref{fig:epg-prompt-test}. 
We add the values of test pass rate in numbers or signs and show EPG in color. 
Empty cells denote that the search halts due to OpenAI exceptions, such as \texttt{APIError}.\footnote{~\url{https://platform.openai.com/docs/guides/error-codes/python-library-error-types}}
\ag{In addition, if the framework halts before max programs attempts (light-blue cells with a~``-''), it is due to the input length limit of the underlying LLM $p_\text{debug}$, i.e., the generated code is too long and does not fit as input to the LLM.}

Some problems are solved with all prompts, while other problems are solved with only a subset of prompts, solved partially, or not solved at all. 
A number of problems are solved with all or the majority of prompts in both languages, such as basement, fizz-buzz, paired-digits, and twitter.
Other problems pass all tests in only one of the languages, such as luhn, vector-distance, fuel-cost, or substitution-cipher. 
Most of the solved problems are generated as the first draft or within 1--2 debug steps. 
However, some problems pass 90\% of test cases at the fifth step, such as substitution-cipher in Python with prompts \ref{prompt-4} and \ref{prompt-8} or shopping-list in C++ with prompts \ref{prompt-0}, \ref{prompt-1}, \ref{prompt-5} and \ref{prompt-7}. 
These runs are likely to be updated with the fully correct programs in the following several steps, according to the results in section 5.1, but the experiments
are stopped for the fairness of inter-prompt comparison.
\ag{Alternatively, conducting the prompt engineering experiment with 1000 max programs would have shown what prompts are beneficial for solving the problems in the long run and can be interesting for future work.}

The most interesting cases concern the problems that are solved only with LLM bug summaries or only with static prompts. 
For example, the gcd problem is solved only with prompts \ref{prompt-6}--\ref{prompt-10} in C++ and is not solved with either of \ref{prompt-0}--\ref{prompt-5}. 
A similar result is obtained for spin-words and coin-sums in C++.
In Python, we observe only the cases where solutions are obtained with static prompts and are not obtained with GPT-assisted prompts, e.g., for find-pair, camel-case. In addition, several prompts work well from both S and M categories as for gcd in Python. 

\begin{mdframed}[style=mystyle]
\vspace*{-1mm}
\noindent
\textbf{Answer to RQ2.} 
Program synthesis in C++ with \method{} achieves better performance in the repair-only setting with both GPT-assisted prompts that summarize bugs in code and static templates which describe failing I/O cases. 
The best-performing C++ instruction is obtained with GPT-3 for text completion that contains the word ``obviously''.
Results differ for PSB2 solutions in Python: the static prompt template \ref{prompt-0} results in the best performance. 
Overall, \method{} performance is stable with different debugging prompts submitted to Codex-edit.
\end{mdframed}
\vspace*{-2mm}

\subsection{Threats to Validity}
\label{sec:threats}

External threats to validity concern \method{} performance on different benchmarks and the use of other language models than the tested ones. 
Specifically, PSB2 contains competitive programming tasks which require smaller functions to be generated than production-scale software.
We plan to extend our experiments in future work to explore the generalizability of results to other benchmarks.

Internal threats relate to the implementation.
We use PSB2, which has corner case tests in the training set and test regular cases in the test set. 
To ensure a fair comparison with other studies on PSB2, we evaluate and report results on the provided test set of PSB2 which risks that the synthesized programs do not pass some of the training cases. 
Large models for code editing and text completion used in this study are nondeterministic, which impacts results. 
Due to prohibitive model inference costs, each experiment was only run once.
However, our temperature sampling procedure described in section \ref{sec:synth} reduces this stochasticity significantly, especially for low-EPG results.
As with other language models, Codex is a black-box model and may generate malicious code~\cite{pearce2021:asleep}. 
\ag{The Codex model was pre-trained on an unbalanced dataset across programming languages~\cite{chen2021:evaluating}. Thus, the results can be skewed towards high performance in popular programming languages.}

\section{Conclusion}
\label{sec:conclusion}
In this study, we propose the \method{} framework to solve the challenge of fully autonomous programming. 
We augment the program synthesis procedure based on the large language models for code generation from templates and textual instructions with the repair block. 
The repair block consists of the tree search across the program candidates generated by a large language model for code.
The LLM used for code repair takes imperfect program candidates and instructions for their improvement as prompts. 
The instructions are obtained from both static templates with failing test case descriptions and templates with auto-generated bug summaries by a text completion language model. 
We explore 11 prompting strategies and the repair-replace trade-off of updating the draft program.

\head{Contributions}
We test \method{} with the Codex-edit as the model for draft program synthesis and debugging in Python and C++ on the PSB2 benchmark. 
In our experiments, \method{} outperforms the PushGP baseline and achieves the state-of-the-art result with 19 solved problems out of 25. 
It requires under 1000 program executions to solve them, in stark contrast to billions\footnote{A problem is considered "solved" by PushGP if at least 1 of 100 runs, each with a limit of 60 million programs, was successful.} of executions in PushGP, making it feasible in the areas with costly testing, such as robotics.
Investigation of the repair-replace trade-off shows that \method{} with tree arity of 10 outperforms both the replace-only strategy and the repair-only approach. 
Our prompt engineering study shows that bug summaries generated with ``confidence indicators'', such as ``obviously'', improve the performance of \method{} during C++ code synthesis. 
Overall, our framework shows low performance variability with different prompts, which indicates its robustness.%

\head{Future work}
To study the generalizability of the \method{} framework, we plan to expand the experiments to the competitive programming dataset of AlphaCode~\cite{li2022:competitionlevel} and QuixBugs~\cite{lin2017:quixbugs}, as well as experimenting with ranking strategies, such as lexicase selection.

\vspace*{-1mm}
\section*{Data Availability}
The code and results are made available via Zenodo.\footnote{~\url{https://doi.org/10.5281/zenodo.7837282}} Note that OpenAI discontinued the Codex API on March 23, 2023, and suggests using the GPT-3.5-Turbo API instead.

\begin{acks}
The work presented in this paper was supported by the European Commission through Horizon 2020 grant 812882, 
and by the Research Council of Norway through the secureIT project (\#288787).
The empirical evaluation made use of the Experimental Infrastructure for Exploration of Exascale Computing (eX3), 
supported by the Research Council of Norway through grant \#270053.
\end{acks}

\printbibliography

\end{document}